\documentclass[%
reprint,
superscriptaddress,
noeprint,
amsmath,amssymb,
longbibliography,
aps,
prx,
]{revtex4-2}
\usepackage{mathtools}
\usepackage{dsfont}
\usepackage{physics,color,comment}
\usepackage{wasysym}
\usepackage{graphicx}
\usepackage{dcolumn}
\usepackage{bm}
\usepackage{hyperref}
\usepackage[capitalize]{cleveref}
\usepackage[mathlines]{lineno}
\usepackage{algorithm} 
\usepackage{algorithmic}
\usepackage{amsthm}
\usepackage{anyfontsize} 

\newtheorem{lem}{Lemma}

\newcommand{\E}[1]{\mathds{E}\left[ #1 \right]}
\newcommand{\V}[1]{\mathds{V}\left[ #1 \right]}
\def \todo #1{\textcolor{red}{#1}}
\def \mc #1{\mathcal{#1}}

\newcommand{\dlangle}{\ensuremath{\langle\!\langle}}
\newcommand{\drangle}{\ensuremath{\rangle\!\rangle}}
\newcommand{\dket}[1]{\lvert #1 \drangle}
\newcommand{\dbra}[1]{\dlangle #1 \rvert}

\makeatletter
\def\dketbra#1{\def\tempa{#1}\futurelet\next\dketbra@i}
\def\dketbra@i{\ifx\next\bgroup\expandafter\dketbra@ii\else\expandafter\dketbra@end\fi}
\def\dketbra@ii#1{\ensuremath{\dket{\tempa}\!\dbra{#1}}}
\def\dketbra@end{\ensuremath{\dket{\tempa}\!\dbra{\tempa}}}
\makeatother

\makeatletter
\def\dbraket#1{\def\tempa{#1}\futurelet\next\dbraket@i}
\def\dbraket@i{\ifx\next\bgroup\expandafter\dbraket@ii\else\expandafter\dbraket@end\fi}
\def\dbraket@ii#1{\ensuremath{\dbra{\tempa}#1 \drangle}}
\def\dbraket@end{\ensuremath{\dbra{\tempa}\tempa\drangle}}
\makeatother

\newcommand\Tstrut{\rule{0pt}{2ex}}       

\begin{document}

\preprint{APS/123-QED}

\title{Independent State and Measurement Characterization for Quantum Computers}

\author{Junan Lin}
\affiliation{Institute for Quantum Computing and Department of Physics and Astronomy, University of Waterloo, Waterloo, Ontario, Canada, N2L 3G1}
\author{Joel J. Wallman}
\affiliation{Institute for Quantum Computing and Department of Applied Mathematics, University of Waterloo, Waterloo, Ontario, Canada, N2L 3G1}
\affiliation{Quantum Benchmark Inc., Kitchener, Ontario, Canada, N2H 4C3}
\author{Ian Hincks}
\affiliation{Quantum Benchmark Inc., Kitchener, Ontario, Canada, N2H 4C3}
\author{Raymond Laflamme}
\affiliation{Institute for Quantum Computing and Department of Physics and Astronomy, University of Waterloo, Waterloo, Ontario, Canada, N2L 3G1}
\affiliation{Perimeter Institute for Theoretical Physics, Waterloo, Ontario, Canada, N2L 2Y5}

\date{\today}

\begin{abstract}
		Correctly characterizing state preparation and measurement (SPAM) processes is a necessary step towards building reliable quantum processing units (QPUs).
		In this work, we discuss the subtleties behind separately measuring SPAM errors.
		We propose a protocol that can separately estimate SPAM errors, in the case where quantum gates are ideal. 
		In the case where the quantum gates are imperfect, we derive bounds on the estimated SPAM error rates, based on gate error measures which can be estimated independently of SPAM processes.
		Our method shows that the gauge ambiguity in characterizing SPAM operations can be resolved, by assuming that there exists one qubit whose initial state is uncorrelated with other qubits in a QPU.
		We test the protocol on a publicly available five-qubit QPU and demonstrate its validity by comparing our results with simulations.
		
	\end{abstract}
	
	\maketitle
	
	\section{Introduction}
	Successfully operating quantum processing units (QPUs) requires sufficiently low error rates.
	Protocols that accurately characterize error rates in different components of a QPU are necessary for testing its quality.
	While there exists many well-developed methods that characterize errors of quantum gates such as quantum process tomography~\cite{Hradil1997quantum,Lvovsky2009continuous,blume2010optimal} and (variants of) randomized benchmarking~\cite{knill2008randomized,magesan2011scalable,magesan2012efficient,carignan-dugas2015characterizing,sheldon2016characterizing,cross2016scalable}, fewer have focused on studying state preparation and measurement (SPAM) errors which can be on the same order as (and sometimes surpass) gate errors in some current QPUs.
	For example, the combined SPAM error in current superconducting transmon qubit systems has been reported to range from 0.8\% to 2\%~\cite{walter2017rapid}, while one- and two-qubit gates may achieve fidelities over $99.9\%$ and $99\%$, respectively~\cite{barends2014nature}.
	The requirement to repeatedly prepare qubits in well-defined initial states and perform syndrome measurements in quantum error correcting codes also puts SPAM operations on the same level of importance as gate operations.
	
	Separately characterizing SPAM is not a straightforward task.
	Conventional approaches, such as quantum state tomography~\cite{Hradil1997quantum} or detector tomography~\cite{fiurasek2001maximum,mogilevtsev2013data,Keith2018}, rely on the existence of some ideal set of measurements or probe states to determine the other.
	Gate-set tomography~\cite{Blume-Kohout2013robust} avoids such unrealistic assumptions by simultaneously determining all state, gate and measurement operators.
	However, such a general treatment can only provide estimates up a gauge transformation~\cite{Jackson2015detecting,Lin2019}, which can alter the relative strength between preparation and measurement errors.
	In this paper, we approach this problem from a new perspective, in view of the above issues.
	After illustrating the problem caused by the gauge freedom and a sufficient assumption to eliminate it, we provide a simple protocol from which the SPAM operators can be separately determined.
	We then derive upper and lower bounds on the estimated parameters in the case of non-ideal quantum gates, based on an error metric that can be estimated independently of SPAM, resolving the self-consistency problem.
	To make the protocol concrete, we performed it on a publicly available five-qubit QPU and obtained consistent results with a simulation.
	Our method provides new insights into the problem of SPAM characterization, and is valuable to validating QPUs.
	Moreover, it complements the many existing protocols that measure gate errors.
	
	\section{SPAM characterization and gauge ambiguity}
	
	The ideal operations on a QPU generally include initializing the qubits in a state described by a density operator $\rho$, applying an arbitrary sequence of unitary gates, and making a final measurement described by a $k$-outcome positive operator-valued measure (POVM) $M=\{M_1, \dots, M_k\}$.
	We will assume here that the state which the QPU can be initialized to is unique.
	The implementation of each of these operations is imperfect due to a variety of noise processes.
	Noisy implementations of operations are denoted with an overset $\sim$ so that, for example, $\tilde{\rho}$ is the noisy implementation of $\rho$.
	To avoid overcrowding the text, we do not put additional $\sim$'s on \emph{parameters} describing any operator: their meaning can usually be understood from context, and additional special notations on parameters will be defined prior to being used.
		
	Denoting the $N$-qubit Pauli basis as $\mathbb{P}^N = \{I,X,Y,Z\}^{\otimes N}$, we can uniquely write
	\begin{equation}\label{eqn_SPAM_definition}
	\begin{gathered}
	\tilde{\rho} = \sum_{P \in \mathbb{P}^N} 2^{-N} \Tr[P^\dagger \tilde{\rho}] P \coloneqq \sum_{P \in \mathbb{P}^N}  2^{-N} s_{P} P,\\
	\tilde{M}_{i} = \sum_{P \in \mathbb{P}^N}  2^{-N} \Tr[P^\dagger \tilde{M}_{i}] P \coloneqq \sum_{P \in \mathbb{P}^N}  2^{-N} m_{P,i} P.
	\end{gathered}
	\end{equation}
	In the Pauli-Liouville representation, $\tilde{\rho}$ is represented as a $4^N \times 1$ vector $\dket{\tilde{\rho}}$ with components $2^{-N/2} s_{P}$~\footnote{$2^{-N/2}$ serves as a normalization factor.}, and similarly $\tilde{M}_i$ as $\dket{\tilde{M}_i}$.
	A linear map $\mc{G}$ is represented by a $4^N \times 4^N$ matrix $\Phi_{\mc{G}}$ with elements
	\begin{equation}\label{eqn_Liouville_map}
	(\Phi_{\mc{G}})_{P, Q} = 2^{-N} \Tr[P\ \mc{G}(Q)].
	\end{equation}
	$\Phi$ is called the Pauli transfer matrix, or PTM.
	In this picture, the result of mapping $\tilde{\mc{G}}$ to a state $\tilde{\rho}$ is given by a matrix multiplication: $\dket{\tilde{\mc{G}}(\tilde{\rho})} = \Phi_{\tilde{\mc{G}}} \dket{\tilde{\rho}}$.
	The probability $p(\tilde{\rho}, \tilde{\mc{G}}, \tilde{M})$ of an outcome corresponding to a POVM element $\tilde{M}$ given an input state $\dket{\tilde{\mc{G}}(\tilde{\rho})}$ can be computed by the inner product via $p(\tilde{\rho}, \tilde{\mc{G}}, \tilde{M}_i) = \dbra{\tilde{M}_i} \Phi_{\tilde{\mc{G}}} \dket{\tilde{\rho}}$.
	
	We now give an operational definition of what ``SPAM errors'' and ``SPAM error rates'' mean.
	Consider the experiment where one prepares the initial state and performs a measurement.
	The ideal and actual probabilities of obtaining outcome $i$ are $\dbraket{M_i}{\rho}$ and $\dbraket{\tilde{M}_i}{\tilde{\rho}}$ respectively.
	We thus define the SPAM error to be the difference between these probabilities for this experiment, that is, as the \textit{vector} $\boldsymbol{\delta}_{\text{SPAM}}$ with components
	\begin{equation}\label{SPAM-def}
	\boldsymbol{\delta}_{\text{SPAM},i} (\tilde{\rho}, \tilde{M}) \coloneqq  \dbraket{M_i}{\rho} - \dbraket{\tilde{M}_i}{\tilde{\rho}}.
	\end{equation}
	Next, the state preparation (SP) error vector $\boldsymbol{\delta}_{\text{SP}}$ and the measurement (M) error vector $\boldsymbol{\delta}_{\text{M}}$ are defined by $\boldsymbol{\delta}_{\text{SP},i} = \boldsymbol{\delta}_{\text{SPAM},i}(\tilde{\rho}, M)$ and $\boldsymbol{\delta}_{\text{M},i} = \boldsymbol{\delta}_{\text{SPAM},i}(\rho, \tilde{M})$, i.e., the SPAM error vector with the measurement/state preparation operators replaced by their ideal versions, respectively.
	
	For a single qubit with a two-outcome measurement, we can always write $\boldsymbol{\delta}_{\text{SPAM}} \coloneqq (1-\epsilon, \epsilon)^T$.
	In the usual case where $\rho = M_0 = \ketbra{0}{0}$, $\epsilon$ corresponds to the probability of returning an outcome $1$ when measuring $\rho$, which we will refer to as the SPAM \emph{error rate}.
	Then, $\boldsymbol{\delta}_{\text{SP}}$ and $\boldsymbol{\delta}_{\text{M}}$ are also each characterized by a single parameter, which we will refer to as the SP-error rate $\epsilon_{\text{SP}}$, and the M-error rate $\epsilon_{\text{M}}$, respectively.
	
	Ideally, one would like to obtain a full, unique description of $\tilde{\rho}$, $\tilde{M}$, and all possible control operations $\tilde{\mc{G}}_j$.
	Unfortunately, this is impossible due to a gauge freedom~\cite{Blume-Kohout2013robust,Jackson2015detecting,Lin2019}.
	In reality, these operators are hidden and we can only infer their values from probabilities based on the Born rule.
	It turns out that the choice of $(\tilde{\rho}, \tilde{\mc{G}}_j, \tilde{M})$ given a list of $p(\tilde{\rho}, \tilde{\mc{G}}_j, \tilde{M})$ is non-unique: they are related by a ``gauge transformation''
	\begin{equation}
	\dket{\tilde{\rho}} \rightarrow B\dket{\tilde{\rho}},\ \dbra{\tilde{M_i}} \rightarrow \dbra{\tilde{M_i}} B^{-1},\ \Phi_{\tilde{\mc{G}}_j} \rightarrow B \Phi_{\tilde{\mc{G}}_j} B^{-1},
	\end{equation}
	where $B$ is an invertible matrix.
	This preserves all outcome probabilities when applied to all elements simultaneously, making the transformed set equally valid as the original set.
	On the other hand, most quality metrics for \textit{individual components} (such as $\epsilon_{\text{SP}}$, $\epsilon_{\text{M}}$, or gate error rates) are \textit{not} gauge-invariant.
	Since separate components in a QPU often require individual calibration in reality, having non-unique metrics is problematic because it becomes unclear whether an operation has improved (e.g., due to a change in control parameter) or not.
	
	Next, we consider the weaker question of SPAM characterization, which boils down to estimating $s_P$ and $m_{P,i}$.
	Previous studies on quantum state and detector tomography showed that $\tilde{\rho}$ or $\tilde{M}$ can be determined if the other is fully known.
	If we assume both to be in the most general form (satisfying only the physicality constraints that $\tilde{\rho}$ is a density matrix and $\tilde{M}$ is a POVM), but allow an arbitrary set of known, unitary gates $\mc{G}_j$, can we learn either $\tilde{\rho}$ or $\tilde{M}$?
	Interestingly, the answer is still negative.
	In particular, since a unitary gate $\mc{G}_j$ is trace-preserving and unital, it can be parametrized by
	\begin{equation}\label{eqn_unitary_PTM}
	\Phi_{\mc{G}_j} = \begin{pmatrix}
	1 & 0\\
	0 & \phi_j
	\end{pmatrix},
	\end{equation}
	where $\phi_j$ is a block matrix with components $\phi_{P,Q,j}$.
	All outcome probabilities are thus in the form
	\begin{equation}\label{eqn_gauge_prob}
	p(\tilde{\rho}, \mc{G}_j, \tilde{M}_i) = 2^{-N} (m_{I^{\otimes N},i} + \sum_{P,Q \in \mathds{P}^{N} \setminus \{I^{\otimes N}\}} \phi_{P,Q,j} s_Q m_{P,i} ),
	\end{equation}
	where we followed the definitions in \cref{eqn_SPAM_definition}~\footnote{note that $s_{I^{\otimes N}} = 1$ by the unit trace constraint.}.
	Among all such equations which can be constructed, $s_Q$ and $m_{P,i}$ always appear in a product form and cannot be separately solved for, assuming that $\phi$ only consists of constants.
	A gauge transformation (named ``blame gauge'' in \cite{Jackson2015detecting}) of the form
	\begin{equation}\label{eqn_gauge_trans}
	s_Q \rightarrow x s_Q,\ m_{P,i} \rightarrow m_{P,i} / x
	\end{equation}
	in the second term for some real number $x$ will keep the equations unaltered~\footnote{Importantly, it also keeps the gate intact, because the matrix $B$ here commutes with $\Phi_{\mc{G}_j}$ for all unitary gates $\mc{G}_{j}$}.
	While this transformation also needs to maintain the physicality constraints on $\tilde{\rho}$ and $\tilde{M}$, it is valid for most experimentally relevant cases~\cite{Lin2019}.
	Therefore, in addition to assuming ideal gates, one needs further assumptions about the structure of SPAM elements, and needs to design an effective operation that breaks this symmetry.
	Below, we will state a sufficient assumption, and present a protocol that achieves this by engineering $\Phi$ to depend upon the SPAM coefficients.

	\section{Protocol assuming ideal gates}
	
	To develop a straightforward protocol, we engineer simplified effective SPAM operators based on an averaging technique in~\cite{wallman2016noise}, by removing undesired components in $\tilde{\rho}$ and $\tilde{M}$.
	From now on, the qubit or system of qubits whose SPAM operators we would like to know will be called the target qubit (system), and we will use the subscript $t$ to indicate parameters of the target qubit (system).
	For now we assume all quantum gates to be ideal, and will relax this later.
	Consider a single qubit initialized to $\tilde{\rho}$ and has a two-outcome POVM $\tilde{M} = \{\tilde{M}_0,\tilde{M}_1 = I - \tilde{M}_0\}$, parametrized by
	\begin{equation}\label{eqn_1q-param}
	\begin{split}
	\dket{\tilde{\rho}} =  \frac{(1, s_X, s_Y, s_Z)^T}{\sqrt{2}},\ \dbra{\tilde{M}_0} = \frac{(m_I, m_X, m_Y, m_Z)}{\sqrt{2}}
	\end{split}
	\end{equation}
	
	\begin{figure*}[ht]
	\centering
	\includegraphics[width=1.5\columnwidth]{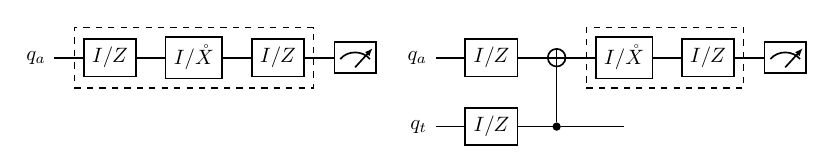}
	\caption{Circuits for determining the coefficient $s_{Z,t}$ on the target qubit $q_t$, assuming ideal gates. The one on the left/right gives $\alpha_a$ and $\beta_t$, respectively.
	Combinations of SPAM averaging gates are run as separate experiments (the outcome is flipped classically when the first M-averaging gate is $X$, indicated by an overhead circle).
	Adjacent single-qubit gates (grouped by the dashed box) are logically compiled to a single gate when running the circuits.}
	\label{fig_1-qubit}
    \end{figure*}
    
\begin{figure}[ht]
	\centering
	\includegraphics[width=1.0\columnwidth]{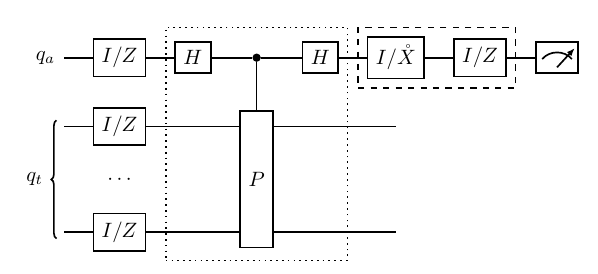}
	\caption{The circuit for estimating $\beta_{P,t}$ of an $N$-qubit system $q_t$, assuming ideal gates. The dotted box indicate the propagating cycle $\mc{U}_{P}$.}
	\label{fig_full_circuit}
    \end{figure}

	We will assume that the ideal state and measurement are $\rho = M_0 = \ketbra{0}{0}$, corresponding to $s_X = s_Y = m_X = m_Y = 0$, and $s_Z = m_Z = 1$ in \cref{eqn_1q-param}.
	In reality these parameters deviate from the ideal, but we can use the following technique to eliminate some undesired components.
	By linearity of quantum operations and probabilities, applying two Pauli gates from the set $\{I,Z\}$ immediately after state preparation and before the measurement, and averaging over outputs from all possible circuits would set $s_X = s_Y = m_X = m_Y = 0$ (this is similar to the phase cycling technique commonly used in nuclear magnetic resonance (NMR) spectroscopy to suppress spurious signals~\cite{Levitt2001}).
	To fix $m_I$, we apply an additional gate from $\{I,X\}$ immediately before the measurement, and relabel the outcome when we apply an $X$ so that the outcome $0$ corresponds to the POVM element $\tilde{M}_1$ and \textit{vice versa}.
	We label this with an overhead circle in \cref{fig_1-qubit} and \cref{fig_full_circuit}.
	Averaging the results from these two circuits (with an $I$ or $X$ averaging gate) effectively sets $m_I = 1$.
    Combining the above, the problem is now reduced to finding $s_{Z,t}$ and $m_{Z,t}$ on the target qubit $q_t$.
    
    A few words regarding SPAM averaging shall take place before we proceed.
    While the SPAM operators after averaging deviate from the original ones, they do represent the ones that actually enter the circuit, if SPAM averaging is consistently applied in all future circuits.
    Since our protocol estimates exactly the parameters in this averaged model, they will predict the correct experimental outputs for future circuits as well.
    
	Now, we provide a simple protocol that estimates $s_{Z,t}$ and $m_{Z,t}$.
	We assume that there exists an ancillary qubit $q_a$ which can be prepared and measured \emph{independently}, i.e. is described by two independent but unknown coefficients $s_{Z,a}$ and $m_{Z,a}$ (after applying the same SPAM averaging).
	If we apply a CNOT gate controlled on $q_t$ and targeted on $q_a$ (see \cref{fig_1-qubit}), which we will call $\mc{C}_{t,a}$, the PTM on $q_a$ can be calculated to be a diagonal matrix
	\begin{equation}\label{eqn_effective_op}
	\Phi = \text{diag}(1,1,s_{Z,t}, s_{Z,t}),
	\end{equation}
	which depends upon $s_{Z,t}$ as desired.
	In other words, the entangling CNOT gate \textit{propagates SP parameters} of $q_t$ to $q_a$.
	Because the measurement on $q_a$ is unaffected by the gate, we can learn $s_{Z,t}$ as follows: define $\alpha$ and $\beta$ as the expectation values $\langle \tilde{M}_0 - \tilde{M}_1 \rangle_a$ on $q_a$ in the \textit{absence} and \textit{presence} of the CNOT gate, respectively.
	A direct calculation [see \cref{eqn_1q-param} and \cref{eqn_effective_op}] shows 
	\begin{equation}\label{eqn_alpha_beta}
	\alpha_{a} =  s_{Z,a} m_{Z,a},\ \beta_{t} = s_{Z,t} s_{Z,a} m_{Z,a},
	\end{equation}
	which gives $s_{Z,t} = \beta_{t} / \alpha_{a}$.
	$m_{Z,t}$ can then be determined by a separate experiment that measures $\alpha_{t}$ on $q_t$ with
	\begin{equation}\label{eqn_m_t}
	    m_{Z,t} = \alpha_{t}/s_{Z,t}.
	\end{equation}
	Note that the subscript for $\alpha$ refers to the qubit \emph{being measured}, while for $\beta$ it refers to the qubit whose \emph{parameter is being propagated}.
	The SP- and M-error rates can then be computed as
	\begin{equation}\label{eqn_epsilon}
	\epsilon_{\text{SP},t} = (1-s_{Z,t})/2,\ \epsilon_{\text{M},t} = (1-m_{Z,t})/2.
	\end{equation}

	We can generalize this idea to measuring parameters of an $N$-qubit system, as long as we assume that there exists \textit{one} additional ancilla $q_a$ that can be prepared and measured independently.
	This is thus a sufficient condition for breaking the gauge symmetry.
	Here, $\alpha_a$ is estimated using the same circuit as the one on the left of \cref{fig_1-qubit}.
	To estimate $\beta_{P,t}$ where $P$ labels the Pauli components of the unknown initial state, the CNOT in the one-qubit case is generalized to $\mc{U}_{P} = (\mc{H}\otimes \mc{I})\mc{C}_P(\mc{H}\otimes \mc{I})$, where $\mc{C}_P$ corresponds to the unitary $\ketbra{0}{0} \otimes I + \ketbra{1}{1} \otimes P$, $P = P_1 \otimes \dots \otimes P_N,\ P_i \in \{I,Z\}$, and $\mc{H}$ corresponds to a Hadamard gate on $q_a$.
	$\mc{C}_P$ is controlled on $q_a$ and targeted on the $N$ qubit system (see \cref{fig_full_circuit}).
	As shown in \cref{sec_appen_gate_effect}, the effect on $q_a$ is identical to \cref{eqn_effective_op} with $s_{Z,t}$ replaced by $s_{P,t}$.
	Repeating for all possible $P$'s will fully determine $\rho_t$, allowing one to determine the POVMs by detector tomography.
	This is summarized in \cref{alg-multq}.
    Note that the number of circuits to average over grows exponentially with the size of $q_t$.
    For large systems, one should instead randomly sample from the space of all SPAM averaging gates.
    Additionally, the controlled-$P$ gate requires $\mc{O}(N)$ controlled-Z gates and has depth $\mc{O}(N)$ for the worst case, but can be achieved by two all-to-all M{\o}lmer-S{\o}rensen gates~\cite{Sorensen1999,Sorensen2000}.	
    
    We would also like to point out that small modifications to \cref{alg-multq} would allow one to obtain \emph{all} components of the SPAM operators in principle.
    For example, in the one-qubit case, one could average over $\{I,X\}$ instead of $\{I,Z\}$ to obtain $\dket{\tilde{\rho}} \sim (1, s_{X,t}, 0,0)^T$.
    Then performing a $Y(-\pi/2)$ rotation would result in $\dket{\tilde{\rho}} \sim (1, 0, 0, s_{X,t})^T$, so that $s_X$ can then be determined in exactly the same way as $s_{Z,t}$ before.
    But this is unnecessary if SPAM averaging is consistently applied for all future circuits, as we discussed previously.

	\begin{algorithm}[H]
      \caption{Estimating the SP- and M-operators of an $N$-qubit system, assuming ideal gates}
      \label{alg-multq}
       \begin{algorithmic}[1]
       \STATE Choose an ancillary qubit $q_a$
       \STATE Measure $q_a$ (see left of \cref{fig_1-qubit}) and record the result $\alpha_{a}$
       \FOR{each $P$ in $\{I,Z\}^{\otimes N}$}
       \STATE Apply the circuits in \cref{fig_full_circuit} with gate $P$, measure $q_a$ and record the result $\beta_{P,t}$
       \STATE Calculate $s_{P,t} = \beta_{P,t}/\alpha_{a}$
       \ENDFOR
       \STATE Apply detector tomography on $q_t$ to estimate $m_{P,j}$ for each POVM element $\tilde{M}_j$, where $j=1,...,k$ for a $k$-outcome POVM.
       \end{algorithmic}
    \end{algorithm}

	\section{Protocol with imperfect gates}
	
	We now take into account gate imperfections.
	The effects of gate errors must be treated in a way that does not rely on any prior information on SPAM, since they are assumed unknown.
	This prohibits using protocols like process tomography to extract the full effect of the gate in question, and substitute to replace $\Phi$ in \cref{eqn_effective_op}.
	Protocols that estimate gate error strengths independently of SPAM offer a solution to the problem.
	Here we utilize the recently proposed cycle benchmarking (CB)~\cite{Erhard2019b} procedure.
	CB estimates the process infidelity of a composite cycle (consisting of a round of the original gates $\tilde{\mc{G}}$ composed with a round of ``dressing'' gates $\tilde{\mc{D}}$), averaged over all Pauli dressing gates, namely 
	\begin{equation}\label{eqn_rCB_defn}
	r_{\text{CB}}(\tilde{\mc{G}},\mc{G}) \coloneqq \sum_{\mc{D} \in \{\mc{I},\mc{X},\mc{Y},\mc{Z}\}^{\otimes N}} 4^{-N} r_p (\tilde{\mc{G}}\tilde{\mc{D}},\mc{G}\mc{D}), 
	\end{equation} 
	where the process infidelity is 
	\begin{equation}\label{eqn_rp_defn}
	r_p(\tilde{\mc{G}},\mc{G}) \coloneqq 1 - 4^{-N} \sum_{P \in \mathds{P}^N} 2^{-N} \Tr[\mc{G}(P)\tilde{\mc{G}}(P)]. 
	\end{equation}
	The figure $r_{\text{CB}}$ is relevant when a quantum computation task is used in conjunction with a noise-tailoring procedure called randomized compiling (RC)~\cite{wallman2016noise}.
	Here, random twirling gates are inserted into the original circuit, such that the logical circuit is preserved.
	Uniformly averaging over all twirling gates turns the error of a composite cycle into a Stochastic Pauli channel $\mc{P}$: $\mc{P}(\rho) = \sum_{P \in \mathds{P}^N} c_P P \rho P^{\dagger}$, where $c_P$'s form a probability distribution.
	The error rate of $\mc{P}$ is then precisely characterized by $r_{\text{CB}}(\tilde{\mc{G}}_j,\mc{G}_j)$~\cite{Erhard2019b}.
	This twirling is exact under the standard assumption that errors on twirling gates are gate-independent (which we assume throughout), and has a relatively small correction when gate dependence is present~\cite{wallman2016noise}.
	For simplicity we also make the standard assumption that one-qubit gates have a one-qubit error channel, however, we conjecture that this can be relaxed.
	

	\begin{table}[ht]
		\centering
		\begin{tabular}{|c|c|c|c|}
			\hline
			$\epsilon_{\text{SP},t}$, lower & $\epsilon_{\text{SP},t}$, upper & $\epsilon_{\text{M},t}$, lower & $\epsilon_{\text{M},t}$, upper \\
			\hline
			
			\scalebox{0.91}{%
			$\displaystyle \frac{1}{2}-\frac{\beta_{t} +2r_{t,a}}{2\alpha_{a}}$} & \scalebox{0.91}{%
            $\displaystyle \frac{1}{2}-\frac{\beta_{t} - 2r_{t,a}}{2\alpha_{a}}$} & \scalebox{0.91}{%
            $\displaystyle \frac{1}{2} -\frac{\alpha_{t} \alpha_{a}}{2\beta_{t} - 4r_{t,a}}$} & \scalebox{0.91}{%
            $\displaystyle \frac{1}{2} -\frac{\alpha_{t} \alpha_{a}}{2\beta_{t} + 4r_{t,a}}$} \Tstrut \\ \hline
		\end{tabular}
		\caption{Upper and lower bounds for one-qubit SPAM error rates [\cref{eqn_epsilon}] on a target qubit $q_t$. $\alpha$ and $\beta$ are defined in \cref{eqn_alpha_beta}. $r_{t,a}$ is shorthand for $r_{\text{CB}}(\tilde{\mc{C}}_{t,a},\mc{C}_{t,a})$.}
		\label{bounds_table}
	\end{table}
	
	Let us now denote the parameters that \textit{would have been obtained with an ideal propagating cycle} with a superscript ic (i.e., ideal cycle).
	These are the actual parameters describing our unknown SPAM operators, which are not affected by the imperfect gates.
	On the other hand, the ones that are actually obtained in experiments will be denoted as normal letters.
	We will show in \cref{sec_appen_beta_bound} that $\beta_{P,t}^{\text{ic}}$ can be bounded using the measured $\beta_{P,t}$ and $r_{\text{CB}}(\tilde{\mc{U}_{P}},\mc{U}_{P})$ as:
	\begin{equation}\label{eqn_beta_bound}
	\beta_{P,t}^{\text{ic}} \in [\beta_{P,t} - 2 r_{\text{CB}}(\tilde{\mc{U}}_{P},\mc{U}_{P}), \beta_{P,t} +2 r_{\text{CB}}(\tilde{\mc{U}}_{P},\mc{U}_{P})],
	\end{equation}
	which holds independently of the dimension of $q_t$.
	Since $s_{P,t} = \beta_{P,t}/\alpha_{a}$, and because $\alpha_{a}$ does not involve gates with unknown effects, we see that 
	\begin{equation}\label{eqn_sP_bound}
	    s_{P,t}^{\text{ic}} \in [\frac{\beta_{P,t} - 2 r_{\text{CB}}(\tilde{\mc{U}}_{P},\mc{U}_{P})}{\alpha_{a}}, \frac{\beta_{P,t} + 2 r_{\text{CB}}(\tilde{\mc{U}}_{P},\mc{U}_{P})}{\alpha_{a}}].
	\end{equation}
	Repeating for all values of $P$ would give a bound on each parameter $s_{P,t}^{\text{ic}}$ of the estimated initial state $\tilde{\rho}$.
	Recall from our previous definition that the $i$-th component of $\boldsymbol{\delta}_{\text{SP}}$ is given by
	\begin{equation}
	    \boldsymbol{\delta}_{\text{SP},i} = \dbra{M_i} (\dket{\rho} - \dket{\tilde{\rho}}) = \sum_{P} m_{P,i} (s_{P}^{\text{ideal}}-s_{P}),
	\end{equation}
	where we used the superscript ``ideal'' to represent the ideal parameters of $\rho$.
	This is a linear function of $s_P$, whose bounds are given by \cref{eqn_sP_bound}.
	Therefore the upper and lower bounds for $\boldsymbol{\delta}_{\text{SP},i}$ can simply be obtained by optimizing each term in the sum, resulting in
	\begin{equation}\label{eqn_del_SP_bound}
	\begin{gathered}
	\boldsymbol{\delta}_{\text{SP},i,\ \text{lower}} = \sum_{P} m_{P,i} s_{P}^{\text{ideal}} - \sum_{P} m_{P,i} s_{\text{sgn}(m_{P,i})},\\
	\boldsymbol{\delta}_{\text{SP},i,\ \text{upper}} = \sum_{P} m_{P,i} s_{P}^{\text{ideal}} - \sum_{P} m_{P,i} s_{-\text{sgn}(m_{P,i})},
	\end{gathered}
	\end{equation}	
	where sgn is the sign function, and we use the shorthand $s_{-}$ and $s_{+}$ to represent the lower and upper bounds in \cref{eqn_sP_bound}.
	
	The bounds for measurement parameters are more complicated, since one would need to perform measurement tomography based on the learned initial state, and different tomography approaches will lead to different bounds.
	But, the principles behind all approaches will be similar.
	Here we demonstrate with the simplest case of a linear inversion (LI) tomography.
	In LI detector tomography, one prepares an informationally-complete set of initial states $\tilde{\rho}_{1}, \dots, \tilde{\rho}_{4^N}$, all of which have been characterized using our procedure by assumption.
	For qubit measurements in the computational basis, the unknown POVM elements will correspond to the outcomes $\ket{0}^{\otimes N} \dots \ket{1}^{\otimes N}$, so there are a total of $2^{N}$ of them.
	Arrange the column vectors $\dket{\tilde{\rho}_{j}}$ into a $4^{N} \times 4^{N}$ matrix $S$.
	Arrange the vectorized POVM elements $\dket{\tilde{M}_i}$ into a $2^{N} \times 4^{N}$ matrix $R$.
	One then measures each basis state $\tilde{\rho}_{i}$ and records the data matrix with components $D_{i,j} = \dbraket{\tilde{M}_{i}}{\tilde{\rho}_{j}}$.
	This gives the matrix relation:
	\begin{equation}
	    R \cdot S = D,
	\end{equation}
	which can be inverted as $R = D \cdot S^{-1}$ to solve for the unknown matrix $R$.
	In the absence of gate error and measurement shot noise, this results in a noiseless reconstruction of the POVM elements $\tilde{M}_{i}$.
	In the presence of gate errors when measuring the states $\tilde{\rho}_{j}$, we have learned from \cref{eqn_del_SP_bound} that each component of the vector $\dket{\tilde{\rho}_{j}}$ is bounded.
	This uncertainty translates into uncertainties in $\tilde{M}_{i}$ through the matrix inverse, $S^{-1}$.
	In this case, each component of the resulted $\dbra{\tilde{M}_{i}}$ is a (potentially highly nonlinear) function of the components of $\dket{\tilde{\rho}_{j}}$.
	Nonetheless, the max and min values of $\dbra{\tilde{M}_{i}}_{k}$ are guaranteed by the extreme value theorem, and can be found using numerical programs such as SCIPY.
	From this, bounds on components of $\boldsymbol{\delta}_{M}$ can be derived in the same way as what we did for $\boldsymbol{\delta}_{SP}$.
	
	The situation becomes particularly simple for one qubit with a two-outcome measurement, along with SPAM averaging.
	In this case there is only one unknown parameter $s_{Z}$ for $\tilde{\rho}$ and another one $m_{Z}$ for $\tilde{M}_{0}$ ($\tilde{M}_{1}$ is fixed by $\tilde{M}_{0}$).
	The bound for $s_{Z}$ is given directly in \cref{eqn_sP_bound}.
	Since $m_{Z}$ and $s_{Z}$ are inversely proportional [see \cref{eqn_m_t}], the maximum of $s_{Z}$ gives the minimum of $m_{Z}$, and \emph{vice versa}.
	We then use \cref{eqn_epsilon} to convert to bounds on the error rates $\epsilon_{\text{SP}}$ and $\epsilon_{\text{M}}$.
	These are summarized in \cref{bounds_table}.
	Intuitively, a smaller gate error corresponds to a narrower range, and the region restores the previous point estimate [\cref{eqn_alpha_beta}] in the limit of perfect gates.
	
	We incorporate gate error effects into \cref{alg-multq} by proposing a simple procedure to separately estimate the single qubit SP- and M-errors in a QPU.
	For each qubit $i$, label it as the ``target'' ($t$) and find an ``ancilla'' ($a$) such that a CNOT gate $\mc{C}_{t,a}$ is allowed by the QPU's connectivity.
	We then run the one-qubit protocol to estimate the SPAM parameters in \cref{bounds_table}.
	Repeatedly identifying each of the $N$ qubits as the ``target'' gives a single-qubit SPAM characterization of the full device.
	This is summarized in \cref{alg-singq}.
    While providing experimentally relevant single-qubit error rates for the full system, the protocol has an overhead that only scales linearly with the system size, making it a practical tool for many scenarios.
    	
    \begin{algorithm}[H]
      \caption{Estimating single qubit SP- and M-error rates on an $N$-qubit device}
      \label{alg-singq}
       \begin{algorithmic}[1]
       \FOR{each of the $N$ qubits}
       \STATE Label it as $q_t$; choose an ancilla $q_a$ where $\mc{C}_{t,a}$ is allowed
       \STATE Run the one-qubit protocol to estimate $\alpha_a$, $\alpha_t$, and $\beta_t$
       \STATE Use cycle benchmarking to estimate $r_{\text{CB}}(\tilde{\mc{C}}_{t,a},\mc{C}_{t,a})$
       \STATE Compute a regional estimate on $\epsilon_{\text{SP},t}$ and $\epsilon_{\text{M},t}$ on $q_t$ according to \cref{bounds_table}.
       \ENDFOR
       \end{algorithmic}
    \end{algorithm}

	We performed the above protocol on a publicly available five-qubit QPU (IBMQ-SANTIAGO~\cite{ibmq-santiago}) to estimate $\epsilon_{\text{SP}}$ and $\epsilon_{\text{M}}$ on each qubit.
	For each target qubit, we chose the ancilla to be the one connected to it with the lowest error CNOT.
	The specification for the experiments can be found in \cref{sec_appen_exp}.
	We used the TrueQ~\cite{Beale2020} software to generate circuits, submit to the IBM-Q server, and perform data analysis.
	The range of possible error rates (i.e., between the upper and lower bounds in \cref{bounds_table}) is shown in shaded regions for each estimated $\epsilon_{\text{SP}}$ or $\epsilon_{\text{M}}$.
	The $95\%$ confidence intervals (CIs) of the upper and lower bounds are shown as error bars, whose derivations can be found in \cref{sec_appen_errors}.
	Any region below zero is discarded due to the physicality constraint that error rates are positive by definition [\cref{eqn_epsilon}].

	\begin{figure}[ht]
	\centering
	\includegraphics[width=1.0\columnwidth]{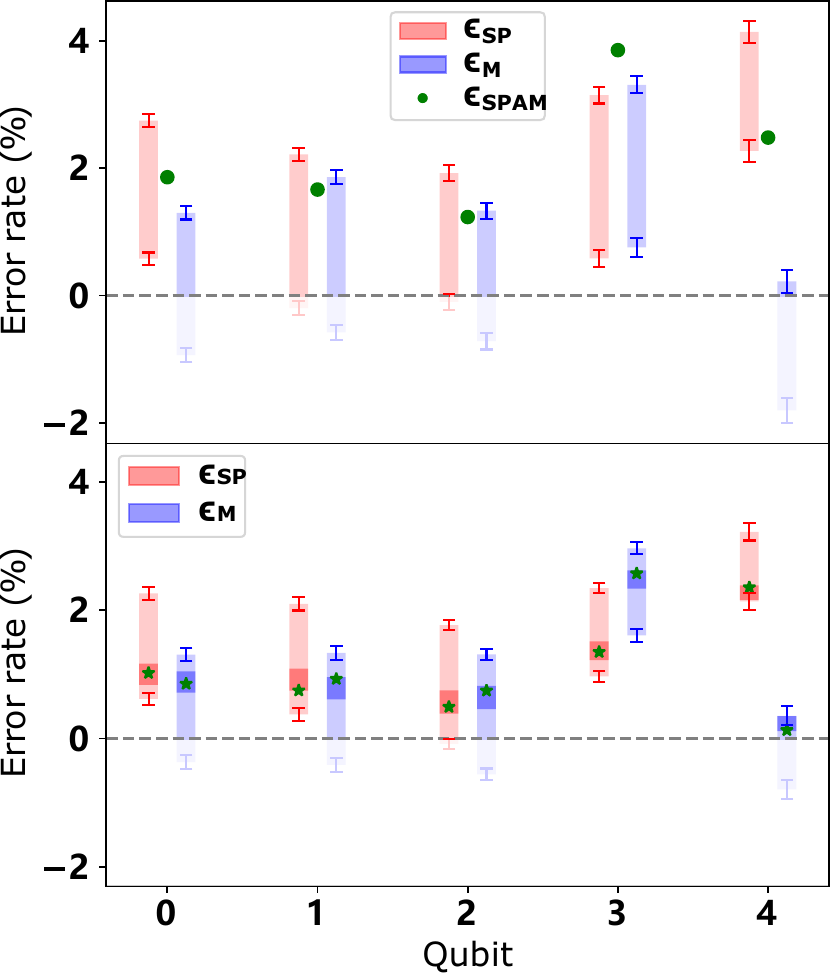}
	\caption{Top: Estimated single qubit state preparation (in red) and measurement (in blue) error rates on IBMQ-SANTIAGO QPU.
	Shaded regions represent the range of error rates consistent with the measured gate error, and error bars are $95\%$ CIs for the endpoints.
	Green dots indicate measured total SPAM error $\epsilon_{\text{SPAM}}$.
	All estimates are cut-off below $0\%$.
	Bottom: A simulation assuming $T_1,\ T_2$ relaxation gate errors on two-qubit gates and ideal one-qubit gates, using the device's specifications.
	Green stars mark the magnitudes of SP- and M-errors used in the simulation, which combine to $\epsilon_{\text{SPAM}}$ on each qubit.
	Darker regions mark the same bounds when all gate times are reduced to $1/5$ of their original values.
	}
	\label{fig-results}
	\end{figure}

    Since IBM-Q does not provide separate SP/M error rates for us to compare with, we simulated the same circuits by manually injecting SPAM errors, in the form of applying a Pauli $X$ gate to the initial $\ket{0}$ state with probability $\epsilon_{\text{SP},i}$ and flipping the (classical) measurement outcome with probability $\epsilon_{\text{M},i}$ independently on each qubit $i$.
    We modeled noisy gates by assuming a simple $T_1 + T_2$ relaxation model, and using the relaxation time and gate time obtained from the provider.
    By manually adjusting the relative magnitudes between $\epsilon_{\text{SP},i}$ and $\epsilon_{\text{M},i}$ (indicated by the green stars) while fixing the total SPAM error to the measured value, we obtained a similar behavior compared to the data, strengthening our claim about noise on the physical device (see the bottom half of \cref{fig-results}).
    Our result shows that state preparation contributes more to the total SPAM error on qubit $4$, while not conclusively on the other qubits.
    Better distinguishability can be achieved if higher quality gates are available, as shown by the darker shaded regions in the bottom of \cref{fig-results}, which represent the same bounds (error bars omitted) in \cref{bounds_table} with all gate times reduced to $1/5$ of their original values.

	\section{Conclusions}
	In this work, we proposed a method to characterize state preparation and measurement errors independently on a QPU.
	In the case where quantum gates are ideal, our method returns the exact state preparation and measurement errors.
	In the case where quantum gates are imperfect, by utilizing randomized compiling and cycle benchmarking techniques, we derived upper and lower bounds for the estimated SPAM errors in terms of gate error rates that can be measured independently of SPAM.
	This resolves the self-consistency issue due to the gauge freedom.
	We demonstrated our protocol on a publicly available QPU and observed consistent results between the data and a computer simulation.
	We believe this protocol can be a valuable tool for benchmarking near-term quantum devices, in complement to the existing protocols that estimate errors on quantum gates.


\begin{acknowledgments}
We acknowledge the use of the IBM Q for this work. The views expressed are those of the authors and do not reflect the official policy or position of IBM or the IBM Q team.
R.L. acknowledges funding from Mike and Ophelia Lazaridis.
J.L. acknowledges fruitful discussions with Tal Mor.
\end{acknowledgments}

\appendix
\section{Calculating the effect of the n-qubit entangling cycle in \cref{fig_full_circuit}}\label{sec_appen_gate_effect}
Here we show that the entangling cycle has the effect of ``propagating'' the desired component from the target qubits to the ancillary qubit.
Specifically, first note that due to the commutation relations between Pauli operators, the only non-zero components of an $N$-qubit state after averaging over $\{I,Z\}$ on each qubit are tensor products of $I$ and $Z$: that is,
\begin{equation}
    \rho_t = \frac{1}{2^N} \sum_{R} s_{R} R,\ R \in \{I,Z\}^{\otimes N}.
\end{equation}
To calculate the effective PTM on $q_a$, we first note that $[P,Q]=0$ for all $P,Q \in \{I,Z\}^{\otimes N}$, and that all elements of $\{I,Z\}^{\otimes N}$ are involutory (meaning that they square to the identity).
First consider the effect of the entangling gate only.
We will consider a particular gate $U = \ketbra{0}{0} \otimes I + \ketbra{1}{1} \otimes T$, where $T$ also belongs to the group of $\{I,Z\}^{\otimes N}$.
The PTM on $q_a$ is given by
\begin{equation}
    (\Phi_{\mc{G}})_{P, Q} = \frac{1}{2^N} \Tr[P\ \mc{G}(Q)]
\end{equation}
where the map $\mc{G}$ is the effect on $q_a$ by first attaching a state $q_t$, applying the controlled-$P$ gate, and tracing out $q_t$.
We can write the term $\mc{G}(Q)$ as
\begin{widetext}
\begin{equation}\label{eqn_appen_gatecalc}
\begin{aligned}
    \mc{G}(Q) &= \Tr_{t}[U(Q\otimes \rho_t)U^{\dagger}]\\
    &= \Tr_{t}[\frac{1}{2} (\ketbra{0}{0} \otimes I + \ketbra{1}{1} \otimes T) (\sum_{R} s_R\  Q\otimes R) (\ketbra{0}{0} \otimes I + \ketbra{1}{1} \otimes T^{\dagger})]
\end{aligned}
\end{equation}
\end{widetext}

\begin{figure*}[ht]
    \centering
    \includegraphics[width=1.4\columnwidth]{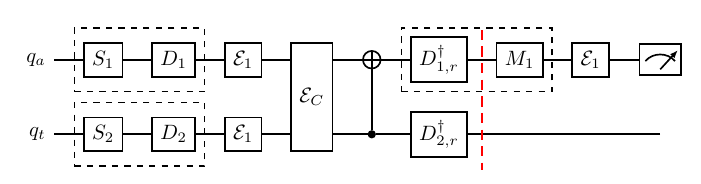}
    \caption{An expanded version of the two-qubit propagating circuits that are actually carried out. The restoring gates are given by $\mc{D}_r^\dagger = \mc{C} \mc{D}^\dagger \mc{C}^\dagger$. Note that SPAM averaging gates are compiled with adjacent dressing gates from randomized compiling (denoted by dashed boxes) and are implemented as one gate, hence there is only one noise channel $\mc{E}_1^{\otimes 2}$. The red dashed line indicates the place where we make the comparison (see text).}
    \label{fig_2q_expanded}
\end{figure*}

Consider two separate cases:
\begin{enumerate}
    \item $Q = I$ or $Q = Z$. Then $Q$ is diagonal and the only non-zero elements are $Q_{00} \coloneqq \bra{0}Q\ket{0}$ and $Q_{11} \coloneqq \bra{1}Q\ket{1}$. Therefore we can simplify \cref{eqn_appen_gatecalc} as
    \begin{equation}
        \begin{aligned}
            &\Tr_{t} [\frac{1}{2^N} \sum_R s_R (\ketbra{0}{0} Q_{00} \otimes R + \ketbra{1}{1} Q_{11} \otimes T R T^{\dagger})]\\
            &= \Tr_{t}[\frac{1}{2^N} \sum_R s_R (\ketbra{0}{0} Q_{00} \otimes R + \ketbra{1}{1} Q_{11} \otimes R)]\\
            & = \Tr_{t}[Q \otimes (\frac{1}{2^N} \sum_R s_R R)]\\
            &=Q
        \end{aligned}
    \end{equation}
    where we used the fact that $T$ commutes with $R$ for all $T,R \in \{I,Z\}^{\otimes N}$ in the first step, that $Q$ is diagonal in the second step, and that density operators have unit trace in the third step.
    Therefore,
    \begin{equation}
        (\Phi_{\mc{G}})_{P, Q} = \frac{1}{2} \Tr[P\ Q] = \delta_{P Q}.
    \end{equation}
    
    \item $Q = X$ or $Q = Y$. Then $Q$ has only off-diagonal elements $Q_{01} \coloneqq \bra{0}Q\ket{1}$ and $Q_{10} \coloneqq \bra{1}Q\ket{0}$. So we can write \cref{eqn_appen_gatecalc} as
    \begin{equation}\label{eqn_appen_gatecalc2}
        \begin{aligned}
            \Tr_{t} [\frac{1}{2^N} \sum_R s_R (\ketbra{0}{1} Q_{01} \otimes R T^{\dagger} + \ketbra{1}{0} Q_{10} \otimes T R)]
        \end{aligned}
    \end{equation}
    Here we use the fact that all elements of $\{I,Z\}^{\otimes N}$ are both Hermitian and involutory, which leads to the relation
    \begin{equation}
        R T^{\dagger} = T R = I^{\otimes N}\ \text{iff} \ T = R.
    \end{equation}
    Since the only element of $\{I,Z\}^{\otimes N}$ with a nonzero trace is $I^{\otimes N}$, these would be the only remaining terms when the partial trace is performed. 
    Keeping only these terms, we further simplify \cref{eqn_appen_gatecalc2} as
    \begin{equation}
        \Tr_{t} [\frac{1}{2^N} \sum_R s_{R=T} (\ketbra{0}{1} Q_{01}+ \ketbra{1}{0} Q_{10}) \otimes I^{\otimes N}]
    \end{equation}
    The term in the parentheses is just $Q$ since $Q = X$ or $Q = Y$ has off-diagonal elements only. Thus we get $\mc{G}(Q) = s_{T} Q$ and
    \begin{equation}
        (\Phi_{\mc{G}})_{P, Q} = \frac{1}{2} \Tr[P\ s_{T} Q] = s_{T} \delta_{P Q}.
    \end{equation}
\end{enumerate}
From the above we get the form of $\Phi_{\mc{G}}$ as:
\begin{equation}
	\Phi_{\mc{G}} = \begin{pmatrix}
	1 & 0 & 0 & 0\\
	0 & s_{T} & 0 & 0\\
	0 & 0 & s_{T} & 0\\
	0 & 0 & 0 & 1
	\end{pmatrix}.
\end{equation}
Finally, the PTM of the a one-qubit Hadamard gate is
\begin{equation}
	\Phi_{H} = \begin{pmatrix}
	1 & 0 & 0 & 0\\
	0 & 0 & 0 & 1\\
	0 & 0 & -1 & 0\\
	0 & 1 & 0 & 0
	\end{pmatrix}.
\end{equation}
The effect of the full cycle is given by the matrix product
\begin{equation}
	\Phi_{H} \Phi_{\mc{G}} \Phi_{H} = \begin{pmatrix}
	1 & 0 & 0 & 0\\
	0 & 1 & 0 & 0\\
	0 & 0 & s_{T} & 0\\
	0 & 0 & 0 & s_{T}
	\end{pmatrix}
\end{equation}
which is identical in form to \cref{eqn_effective_op} in the main text.

\section{Proof of \Cref{eqn_beta_bound}}\label{sec_appen_beta_bound}
In this section we prove \cref{eqn_beta_bound} in the main text.
First, we define the operator 1-norm $\norm{A}_1 \coloneqq \Tr[\sqrt{A^\dagger A}]$, the induced superoperator norm $\norm{\mc{G}}_{1\rightarrow 1} \coloneqq \max\{\norm{\mc{G}(A)}_{1}: \norm{A}_1 \leq 1\}$ and the diamond norm $\norm{\mc{G}}_{\diamond} \coloneqq \norm{\mc{G} \otimes \mc{I}_d}_{1\rightarrow 1}$ for quantum channels~\cite{gilchrist2005distance,watrous2018theory}.
Note that $\norm{\rho}_1 = 1$ for any density matrix $\rho$.
We would like to compare the distance between the final states of the ancillary qubit $q_a$ immediately before the measurement (see \cref{fig_2q_expanded}), under two cases: the imaginary case where the propagation cycle is ideal (denoted as $\rho_{a}^{\text{ic}}$, ``ideal cycle''), and the actual case with imperfect gates (denoted as $\rho_{a}$).

Below we denote a round of CNOT gate as $\mc{C}$, and the channel represented by the round of dressing gate $D_1 \otimes D_2$ as $\mc{D}$.
Under the assumption of gate-independent error on the dressing gates, we will denote this (single-qubit) error channel as $\mc{E}_1$ so that $\tilde{\mc{D}} = \mc{E}_1 \mc{D}$.
The error on the CNOT can be a general channel denoted by $\tilde{\mc{C}} = \mc{C} \mc{E}_C$.
The dressing gates are randomly sampled from the two-qubit Pauli channels $\{\mc{I},\mc{X},\mc{Y},\mc{Z}\}^{\otimes 2}$.
Recall that $\rho_{a}$ denotes the state of the ancilla qubit immediately before measurement.
Further, let us denote the full system immediately before measurement as $\rho_{\text{full}}$.
We then have the following chain of inequalities:
\begin{equation}
    \begin{aligned}
    \norm{\rho_{a,\mc{D}} - \rho_{a,\mc{D}}^{\text{ic}}}_1 &= \norm{\Tr_2[\mc{D}_r^\dagger(\rho_{\text{full}} - \rho_{\text{full}}^{\text{ic}})]}_1\\
    &\leq \norm{\mc{D}_r^\dagger(\rho_{\text{full}} - \rho_{\text{full}}^{\text{ic}})}_1\\
    & \leq \norm{\rho_{\text{full}} - \rho_{\text{full}}^{\text{ic}}}_1\\
    & \leq \norm{\tilde{\mc{C}}\tilde{\mc{D}} - \mc{C}\mc{D}}_{\diamond}
    \end{aligned}
\end{equation}
for each choice of dressing gates $\mc{D}$.
The first inequality is because partial tracing does not increase the trace distance~\cite{Nielsen2010}.
The second inequality is because trace distance is non-increasing upon action of any CPTP map.
The third inequality is from the definition of the diamond norm.

The diamond norm is related to the process infidelity $r_p$ [\cref{eqn_rp_defn} in the main text] by~\cite{wallman2016noise}
\begin{equation}\label{eps and r}
2 r_p(\tilde{\mc{C}}\tilde{\mc{D}}, \mc{C}\mc{D}) \leq \norm{\tilde{\mc{C}}\tilde{\mc{D}} - \mc{C}\mc{D}}_{\diamond} \leq 2 d \sqrt{r_p(\tilde{\mc{C}}\tilde{\mc{D}}, \mc{C}\mc{D})}
\end{equation}
for $d$-dimensional channels.
From our assumption on the error model, $\tilde{\mc{C}}\tilde{\mc{D}} = \mc{C} \mc{E}_C \mc{E}_1^{\otimes 2} \mc{D}$.
Note that for any channel $\mc{E}$ and unitary processes $\mc{U}$ and $\mc{V}$,
\begin{equation}
    r_p(\mc{E},\mc{U}) = r_p(\mc{U}^\dagger \mc{E}, \mc{I}),\ \norm{\mc{U} \mc{E} \mc{V}}_{\diamond} = \norm{\mc{E}}_{\diamond},
\end{equation}
and since both $\norm{\cdot}_\diamond$ and $r_p$ are linear functions in their arguments,
\begin{equation}
\begin{aligned}
    r_{\text{CB}}(\tilde{\mc{C}}, \mc{C}) &= 4^{-2} \sum_{\mc{D} \in \{\mc{I},\mc{X},\mc{Y},\mc{Z}\}^{\otimes 2}} r_p(\tilde{\mc{C}}\tilde{\mc{D}}, \mc{C}\mc{D})\\
    &= 4^{-2} \sum_{\mc{D} \in \{\mc{I},\mc{X},\mc{Y},\mc{Z}\}^{\otimes 2}} r_p(\mc{D}^\dagger\mc{C}^\dagger\tilde{\mc{C}}\tilde{\mc{D}}, \mc{I})\\
    &= 4^{-2} \sum_{\mc{D} \in \{\mc{I},\mc{X},\mc{Y},\mc{Z}\}^{\otimes 2}} r_p(\mc{D}^\dagger \mc{E}_C \mc{E}_1^{\otimes 2} \mc{D},\mc{I})\\
    &= r_p(\mc{P},\mc{I}),
\end{aligned}\label{eqn_rcb_rp}
\end{equation}
and
\begin{equation}
    4^{-2} \sum_{\mc{D} \in \{\mc{I},\mc{X},\mc{Y},\mc{Z}\}^{\otimes 2}} \norm{\tilde{\mc{C}}\tilde{\mc{D}} - \mc{C}\mc{D}}_{\diamond} = \norm{\mc{P} - \mc{I}}_{\diamond} ,
\end{equation}
where $\mc{P}$ is the twirled error channel mentioned in the main text.
After averaging over all Pauli dressing gates (i.e., the circuit is randomly compiled), 
\begin{equation}
\begin{aligned}
    \norm{\rho_{a,\text{RC}} - \rho_{a,\text{RC}}^{\text{ic}}}_1     &\coloneqq 4^{-2} \sum_{\mc{D} \in \{\mc{I},\mc{X},\mc{Y},\mc{Z}\}^{\otimes 2}} \norm{\rho_{a,\mc{D}} - \rho_{a,\mc{D}}^{\text{ic}}}_1 \\
    &\leq 4^{-2} \sum_{\mc{D} \in \{\mc{I},\mc{X},\mc{Y},\mc{Z}\}^{\otimes 2}} \norm{ \tilde{\mc{C}}\tilde{\mc{D}} - \mc{C}\mc{D}}_{\diamond}\\
    &= \norm{\mc{P} - \mc{I}}_{\diamond} \\
    &= 2 r_p(\mc{P},\mc{I}),
\end{aligned}
\end{equation}
where the last equality is because the lower bound of \cref{eps and r} is saturated for a Pauli noise channel.
Combining with \cref{eqn_rcb_rp}, we finally have
\begin{equation}\label{eqn_appen_1norm_rCB}
    \norm{\rho_{a,\text{RC}} - \rho_{a,\text{RC}}^{\text{ic}}}_1 \leq 2 r_{\text{CB}}(\tilde{\mc{C}},\mc{C}).
\end{equation}

On the other hand, $\norm{\rho_{a,\text{RC}} - \rho_{a,\text{RC}}^{\text{ic}}}_1$ is related to $\beta$ and $\beta^{\text{ic}}$ through the Holevo-Helstrom theorem for distinguishing quantum states~\cite{watrous2018theory}.
We quote theorem 3.4 in \cite{watrous2018theory} as the following lemma:
\begin{lem} 
    Let $\rho_1,\ \rho_2$ be density operators. Let $\lambda \in [0,1]$. For an arbitrary two-outcome POVM measurement described by elements $\{M_0, M_1\}$, it holds that
    \begin{equation}
        \lambda \langle M_0, \rho_0 \rangle + (1-\lambda) \langle M_1, \rho_1 \rangle \leq \frac{1}{2} (1 + \norm{\lambda \rho_0 - (1-\lambda)\rho_1}_{1}).
    \end{equation}
\end{lem}
From the definition of $\beta$, we can rewrite it as
\begin{equation}
    \begin{gathered}
    \beta_{\mc{D}} = 2 \langle M_0, \rho_{a,\mc{D}} \rangle - 1 = 1 - 2\langle M_1, \rho_{a,\mc{D}} \rangle \\
    \beta_{\mc{D}}^{\text{ic}} = 2 \langle M_0, \rho_{a,\mc{D}}^{\text{ic}} \rangle - 1 = 1 - 2\langle M_1, \rho_{a,\mc{D}}^{\text{ic}} \rangle
    \end{gathered}
\end{equation}
for each particular $\mc{D}$.
Since the measurement on $q_a$ is unchanged by the propagation cycle, we can apply the above lemma with $\lambda = \frac{1}{2}$ twice: first, using the first definition of $\beta_{\mc{D}}$ and the second definition of $\beta_{\mc{D}}^{\text{ic}}$,
\begin{equation}
    \beta_{\mc{D}} - \beta_{\mc{D}}^{\text{ic}} = 2 (\langle M_0, \rho_{a,\mc{D}} \rangle + \langle M_1, \rho_{a,\mc{D}}^{\text{ic}} \rangle - 1) \leq \norm{\rho_{a,\mc{D}}^{\text{ic}} - \rho_{a,\mc{D}}}_1
\end{equation}
and next, using the second definition of $\beta_{\mc{D}}$ and the first definition of $\beta_{\mc{D}}^{\text{ic}}$,
\begin{equation}
    \beta_{\mc{D}}^{\text{ic}} - \beta_{\mc{D}} = 2 (\langle M_0, \rho_{a,\mc{D}}^{\text{ic}} \rangle + \langle M_1, \rho_{a,\mc{D}} \rangle - 1) \leq \norm{\rho_{a,\mc{D}} - \rho_{a,\mc{D}}^{\text{ic}}}_1
\end{equation}
Combining the above two equations we get:
\begin{equation}
	-\norm{\rho_{a,\mc{D}}^{\text{ic}} - \rho_{a,\mc{D}}}_1 \leq \beta_{\mc{D}}^{\text{ic}} - \beta_{\mc{D}} \leq \norm{\rho_{a,\mc{D}}^{\text{ic}} - \rho_{a,\mc{D}}}_1.
\end{equation}
Thus, combining with \cref{eqn_appen_1norm_rCB} and averaging over all $\mc{D}$'s, we obtain the desired result
\begin{equation}
    \abs{\beta_{\text{RC}}^{\text{ic}} - \beta_{\text{RC}}} \leq 2 r_{\text{CB}}(\tilde{\mc{C}},\mc{C}).
\end{equation}
The proof can be trivially extended to the case where a multi-qubit propagation cycle, $(\mc{H}\otimes \mc{I})\mc{C}_P(\mc{H}\otimes \mc{I})$, is used in place of the CNOT gate, hence \cref{eqn_beta_bound} in the main text.

\section{Calculating uncertainties in estimated parameters}\label{sec_appen_errors}
In this section, we derive the expressions for uncertainties in our experiments.
The directly measured quantities in our scheme are the expectation values $\alpha$ and $\beta$, as well as the infidelity $r_{\text{CB}}$ measured by cycle benchmarking.
Here, we will focus on the uncertainties on $\alpha$ and $\beta$.
The one for $r_{\text{CB}}$ is based on the same ideas but involves more technical details, and we refer to the Supplementary Information of the original paper~\cite{Erhard2019b} for the exact expressions.
In this section only, we will denote the estimated values of the random variables with an overhead tilde, such that the measured value of $\alpha$ is written as $\tilde{\alpha}$.
We will first derive estimators for the expectation value (denoted with $\mathds{E}$) and variance (denoted with $\mathds{V}$) of a desired quantity in the general case, then apply it to the case of $\alpha$ and $\beta$.
Finally we use standard error propagation to obtain the uncertainty on the upper and lower bounds.

The quantity of interest which we try to estimate can generally be described by the following average value:
\begin{equation}\label{eqn_appen_general_sum}
    \lambda = \frac{1}{N} \sum_{i=1}^{N} p_i,
\end{equation}
where the value of $N$ depends on the context.
In our case, $\lambda$ can be $\alpha$ or $\beta$, so the $p_i$'s are expectation values of single qubit measurements.
There are two things to be noted about estimating this quantity: first, $N$ can be very large in general, so that it is sometimes not possible to exhaustively sample all $p_i$'s.
Second, each $p_i$ cannot be measured perfectly because of finite sampling error.
From now on, we will assume that we sample $n$ out of the $N$ elements with or without replacement.
For a particular $n$-element sample $s$, the value of each sampled element $\tilde{p}_{s_i}$ is a random variable which is denoted with a hat.
Furthermore we assume that $\E{\tilde{p}_i} = p_i$, and that the variance $\V{\tilde{p}_i} \coloneqq \sigma_i$ exists and can be estimated using an unbiased estimator $\tilde{\sigma}_i$.

We now derive the estimators of interest.
By a simple counting argument (see, for example, Sec. 2.6 of \cite{Thompson2012}) and the law of total expectation, it is easy to see that 
\begin{equation}
    \E{\frac{1}{n} \sum_{s_i}^{n} \tilde{p}_{s_i}} = \mathds{E}_s \left[\frac{1}{n} \sum_{s_i}^{n} \E{\tilde{p}_{s_i}}\right] = \frac{1}{N} \sum_i p_i,
\end{equation}
where the standard abuse of notation of denoting the sample with subscript (and summation) is used.
This holds true whether we are sampling with or without replacement.
Therefore the estimator 
\begin{equation}\label{eqn_appen_mean_est}
    \frac{1}{n} \sum_{s_i}^{n} \tilde{p}_{s_i}
\end{equation}
is an unbiased estimator for the population mean.
Next, we can use the law of total variance to compute the variance of this estimator.
Note that in the absence of noisy measurements ($\tilde{p}_i = p_i$) the problem reduces to estimating the variance of the sample mean, which has a well-known formula (see Sec. 2.5 and 2.6 of \cite{Thompson2012}) when the sampling is done without replacement:
\begin{equation}
    \mathds{V}_s \left[ \frac{1}{n} \sum_{s_i}^{n} p_{s_i} \right] = \left(1-\frac{n}{N}\right) \frac{\sigma^2}{n}
\end{equation}
or with replacement:
\begin{equation}
    \mathds{V}_s \left[ \frac{1}{n} \sum_{s_i}^{n} p_{s_i} \right] = \frac{N-1}{N} \frac{\sigma^2}{n}
\end{equation}
where $\sigma^2$ is the population variance of $p_i$:
\begin{equation}
    \sigma^2 = \frac{1}{N-1} \sum_{i=1}^{N}  (p_i - \bar{p})^2,\ \bar{p} = \sum_{i=1}^{N} \frac{p_i}{N}.
\end{equation}
By the law of total variance we can extend to account for noisy measurements: for sampling without replacement we have
\begin{align}
    &\V{\frac{1}{n} \sum_{s_i}^{n} \tilde{p}_{s_i}}\\ 
    =& \mathds{V}_s \E{\left.\frac{1}{n} \sum_{s_i}^{n} \tilde{p}_{s_i} \right\vert s} + \mathds{E}_s \V{\left.\frac{1}{n} \sum_{s_i}^{n} \tilde{p}_{s_i} \right\vert s}\\
    =& \mathds{V}_s \left[ \left.\frac{1}{n} \sum_{s_i}^{n} p_{s_i} \right\vert s \right] + \mathds{E}_s \left[ \frac{1}{n^2} \sum_{s_i}^{n} \sigma_i^2 \right]\\
    =&  \left(1-\frac{n}{N}\right) \frac{\sigma^2}{n} + \frac{1}{n} \sum_{i=1}^{N} \frac{\sigma_i^2}{n}\\
    =&  \left(1-\frac{n}{N}\right) \frac{\sigma^2}{n} + \frac{1}{nN} \sum_{i=1}^{N} \sigma_i^2,\label{eqn_appen_variance}
\end{align}
and similarly for with replacement,
\begin{equation}
    \V{\frac{1}{n} \sum_{s_i}^{n} \tilde{p}_{s_i}} = \left(1-\frac{1}{N}\right) \frac{\sigma^2}{n} + \frac{1}{nN} \sum_{i=1}^{N} \sigma_i^2.
\end{equation}
An important implication from the above expression is that, for situations where each $\sigma_i$ is small, or where $N$ is very large, the above expression depends mostly on the spread of the quantities over the set of values (i.e., $\sigma^2$) and only very weakly on $N$. 
A practical example that aligns with our protocol is where $N$ grows exponentially in the number of qubits, and $\sigma_i$ decreases as the square root of measurement ``shots.''
This ensures that randomly sampling from a large population is scalable in practice.

We then need an estimator for $\sigma^2$ and $\frac{1}{N} \sum_{i=1}^{N} \sigma_i^2$.
For the second quantity it is simply $\frac{1}{n} \sum_{i=1}^{n} \tilde{\sigma}_{s_i}^2$, which are variances of each element in the chosen sample.
For the first quantity, it can be shown that the sample variance corrected by the average of $\sigma_{s_i}^2$'s gives an unbiased estimator for $\sigma^2$: i.e.,
\begin{equation}\label{eqn_appen_sigma_est}
    \E{\tilde{s}^2 - \frac{1}{n} \sum_{i}^{n} \tilde{\sigma}_{s_i}^2} = \sigma^2.
\end{equation}
where $\tilde{s}^2$ is the variance for the chosen sample.
To see this, note that $\sigma^2$ is equal to
\begin{align}
    \sigma^2 &= \frac{1}{N-1} (p_i^2 - \frac{2}{N} \sum_{j}^{N} p_i p_j + \frac{1}{N^2} \sum_{jj'} p_{j} p_{j'})\\
    &= \frac{1}{N-1} (\sum_{i}^{N} p_i^2 - \frac{1}{N} \sum_{ij} p_i p_j)\\
    &= \frac{1}{N-1} (\sum_{i}^{N} p_i^2 - \frac{1}{N} (\sum_{i}^{N} p_i^2 + 2 \sum_{i\neq j}^{N} p_i p_j))\\
    &= \frac{1}{N} \sum_{i}^{N} p_i^2 - \frac{2}{N(N-1)} \sum_{i\neq j}^{N} p_i p_j.\label{eqn_appen_sigma^2}
\end{align}
Meanwhile, the expectation value of $\tilde{s}^2$ is
\begin{align}
        \E{\tilde{s}^2} &= \E{\frac{\sum_{i=1}^{n}(\tilde{p}_{s_i} - \frac{1}{n} \sum_{j=1}^{n} \tilde{p}_{s_j})^2}{n-1}} \\
        &= \frac{1}{n-1} \E{\sum_{i}^{n}\tilde{p}_{s_i}^2 - \frac{1}{n}\sum_{i,j}^{n}\tilde{p}_{s_i} \tilde{p}_{s_j}}.\label{eqn_appen_s^2}
\end{align}
By a counting argument, the first term evaluates to
\begin{align}
    \E{\sum_{i}^{n}\tilde{p}_{s_i}^2} &= \frac{n}{N} \sum_{i}^{N} \E{\tilde{p}_{i}^2} = \frac{n}{N} \sum_{i}^{N} p_i^2 + \sigma_i^2.\label{eqn_appen_p^2}
\end{align}
The second term evaluates to
\begin{align}
    &\E{\frac{1}{n}\sum_{i,j}^{n}\tilde{p}_{s_i} \tilde{p}_{s_j}} = \E{\frac{1}{n} (\sum_{i=j}^{n} +\sum_{i\neq j}^{n} ) \tilde{p}_{s_i} \tilde{p}_{s_j}}\\
    = &\frac{1}{N} \sum_{i}^{N} \E{\tilde{p}_i^2} + \frac{2(n-1)}{N(N-1)} \sum_{i \neq j}^{N} \E{\tilde{p}_{i} \tilde{p}_{j}}\\
    = &\frac{1}{N} \sum_{i}^{N} (p_i^2 + \sigma_i^2) + \frac{2(n-1)}{N(N-1)} \sum_{i\neq j}^{N} p_i p_j\label{eqn_appen_pij}
\end{align}
where the last equality is given by independence of $p_i$'s.

Combining \cref{eqn_appen_s^2,eqn_appen_p^2,eqn_appen_pij} and compare with \cref{eqn_appen_sigma^2}, one sees that the corrected estimator [\cref{eqn_appen_sigma_est}] is unbiased.
We can then use this in \cref{eqn_appen_variance} and simplify to get our unbiased estimator for the variance for sampling without replacement:
\begin{align}
    &(\frac{1}{n} - \frac{1}{N}) (\tilde{s}^2 - \frac{1}{n} \sum_{i}^{n} \tilde{\sigma}_{s_i}^2) + \frac{1}{n^2} \sum_{i}^{n} \tilde{\sigma}_{s_i}^2\\
    = &(\frac{1}{n} - \frac{1}{N}) \tilde{s}^2 - (\frac{1}{n^2} - \frac{1}{nN})\sum_{i}^{n}\tilde{\sigma}_{s_i}^2 + \frac{1}{n^2} \sum_{i}^{n} \tilde{\sigma}_{s_i}^2\\
    = &(\frac{1}{n} - \frac{1}{N}) \tilde{s}^2 + \frac{1}{nN} \sum_{i}^{n} \tilde{\sigma}_{s_i}^2,\label{eqn_appen_var_est}
\end{align}
and sampling with replacement:
\begin{align}
    &(\frac{1}{n} - \frac{1}{nN}) (\tilde{s}^2 - \frac{1}{n} \sum_{i}^{n} \tilde{\sigma}_{s_i}^2) + \frac{1}{n^2} \sum_{i}^{n} \tilde{\sigma}_{s_i}^2\\
    = &(\frac{1}{n} - \frac{1}{nN}) \tilde{s}^2 + \frac{1}{n^2 N} \sum_{i}^{n} \tilde{\sigma}_{s_i}^2.\label{eqn_appen_var_est_repl}
\end{align}

\cref{eqn_appen_mean_est} and \cref{eqn_appen_var_est} [or \cref{eqn_appen_var_est_repl}] allow us to write down mean and variance estimators for any quantity in the form of \cref{eqn_appen_general_sum}.
We then use standard (linear-approximated) error propagation to estimate uncertainties on the parameters of interest, i.e., upper and lower bound on the error rates $\epsilon_{\text{SP}}$ and $\epsilon_{\text{M}}$ from Table 1 in the main text.
Each bound is an independent estimate and is a function of the four parameters: $\alpha_t,\ \alpha_a,\ \beta_t$, and $r_{t,a}$.
The uncertainty on individual parameters are independent of each other, so its covariance matrix $\Sigma^p$ is diagonal (where $p$ stands for ``parameter'').
The first order approximation to the covariance matrix of the bounds is given by $\Sigma^b = J \Sigma^p J^T$ where $J$ is the Jacobian.
We then take the diagonal elements of $\Sigma^b$ to be the uncertainties of the bounds.

Finally we mention how each $\sigma_{s_i}^2$ is estimated for our experiment.
Since $\tilde{p}_i$ equals to a binomial variable divided by the sample size $k$, it has mean $p_i$ and variance $\frac{p_i(1-p_i)}{k}$.
Using again the relation $\E{\tilde{p}^2} = p^2 + \sigma^2$, it can be verified that an unbiased estimator for the variance is $\frac{\tilde{p}_i(1-\tilde{p}_i)}{k-1}$.
Because $\tilde{\alpha}_i = 2 \tilde{p}_i-1$, $\sigma_{\alpha,i}^2 = 4 \sigma_{p,i}^2$.
We can express the estimator in terms of $\tilde{\alpha}_i$ as 
\begin{equation}\label{eqn_appen_sigma_alpha}
    \tilde{\sigma}_{\alpha,i}^2 = \frac{(1+\tilde{\alpha_i})(1-\tilde{\alpha_i})}{k-1}.
\end{equation}
The estimator for $\beta$ is identical, except changing the $\tilde{\alpha}_i$ to $\tilde{\beta}_i$ in the above expression.

\section{Specifications for the experiment and simulation}\label{sec_appen_exp}
The presented experiment was performed on the IBMQ-SANTIAGO machine on Jan. 28, 2021.
The specifications for each different type of experiment are summarized as follows:
\begin{enumerate}
    \item Each $\alpha$ was obtained by averaging exhaustively over the four possible cases, corresponding to the cases where the compiled SPAM randomizing gates belong to $\{I,X,Y,Z\}$. 
    Each circuit is sampled with $8192$ measurement shots. 
    \item Each $\beta$ is estimated by averaging over $60$ randomly compiled circuits (with a total of 256).
    Each circuit is sampled with $k=1024$ measurement shots.
    For practical convenience, we performed sampling with replacement because the estimate precision is sufficiently high; if higher precision is desired, one may switch to sampling without replacement or even exhaustive sampling as in estimating $\alpha$. 
    \item The infidelity $r_{t,a}$ for each CNOT was estimated using cycle benchmarking by repeating the CNOT cycle $\{4, 84\}$ times, and averaging over all $16$ Pauli decay strings, each by sampling $30$ random circuits with $128$ shots.
    The specific choice of CNOT gates used in the experiment were: $\mc{C}_{0,1},\ \mc{C}_{1,0},\ \mc{C}_{2,1},\ \mc{C}_{3,2},\ \mc{C}_{4,3}$.
\end{enumerate}

Next we sketch how the simulation was performed.
For each qubit, we individually add a state preparation error to it by replacing the ideal initial state $\ket{0}$ with a density matrix $\rho = \text{diag}(1-\epsilon_{\text{SP}}, \epsilon_{\text{SP}})$.
We add a measurement error by classically flipping the outcome (symmetrically, from 0 to 1 and from 1 to 0) with probability $\epsilon_{\text{M}}$.
Gate errors are simulated using a simple $T_1$, $T_2$ relaxation model: for each clock cycle in the circuit, we apply a noise process to each qubit defined by the following Choi matrix
\begin{equation}
C = \begin{pmatrix}
    1 & 0 & 0 & e^{-t/T_2} \\
    0 & e^{-t/T_1} & 0 & 0 \\
    0 & 0 & 0 & 0 \\
    e^{-t/T_2} & 0 & 0 & 1-e^{-t/T_1}
    \end{pmatrix}.
\end{equation}
The $T_1$ and $T_2$ relaxation times are obtained from the provider and are tabulated in \cref{table_appen_T1T2}.
Single-qubit gates all have the same (35.6ns) gate time, according to the provider.
The gate time of the CNOT gates used in the experiments are tabulated in \cref{table_appen_gatetime}.

\begin{table}[ht]
\begin{tabular}{|c|c|c|c|c|c|}
\hline
Qubit & 0 & 1 & 2 & 3 & 4 \\ \hline
$T_1$ ($\mu$s) & 75.9 & 134.667 & 120.21 & 137.32 & 100.68\\ \hline
$T_2$ ($\mu$s) & 140.18 & 96.34 & 87.25 & 94.1 & 133.12 \\ \hline
\end{tabular}
\caption{$T_1$ and $T_2$ for individual qubits on IBMQ-SANTIAGO.}\label{table_appen_T1T2}
\end{table}

\begin{table}[ht]
\begin{tabular}{|c|c|c|c|c|c|}
\hline
Gate & $\mc{C}_{0,1}$ & $\mc{C}_{1,0}$ & $\mc{C}_{2,1}$ & $\mc{C}_{3,2}$ & $\mc{C}_{4,3}$ \\ \hline
Gate time (ns) & 526.2 &561.7 &568.9 &412.4 &341.3 \\ \hline
\end{tabular}
\caption{Gate time of each CNOT gate used in the experiment.}\label{table_appen_gatetime}
\end{table}

\bibliography{SPAM-sep}

\end{document}